\documentclass[twocolumn,showpacs,preprintnumbers,amsmath,amssymb]{revtex4}

\usepackage{graphicx}
\usepackage{dcolumn}
\usepackage{bm}
\begin{document}

\preprint{APS/123-QED}

\title{Multiphonon emission model of spin-dependent exciton formation in organic semiconductors}

\author{M. Wohlgenannt}
\email{markus-wohlgenannt@uiowa.edu}
\affiliation{Department of Physics and Astronomy, The University of Iowa, Iowa City, IA 52242-1479}%

\date{\today}

\begin{abstract}
The maximum efficiency in organic light-emitting diodes (OLEDs) depends on the ratio, $r=k_S/k_T$, where $k_S$ ($k_T$) is the singlet (triplet) exciton formation rate. Several recent experiments found that r increases with increasing oligomer length from a value $r \approx 1$ in monomers and short oligomers. Here, we model exciton formation as a multi-phonon emission process. Our model is based on two assertions: (i) More phonons are emitted in triplet formation than in singlet formation. (ii) The Huang-Rhys parameter for this phonon emission is smaller in long oligomers than in short ones. We justify these assertions based on recent experimental and theoretical data.
\end{abstract}

\pacs{73.50.Jt,73.50.Gr,78.60.Fi}
\maketitle

\section{Introduction}

The maximum possible internal quantum efficiency, $\eta_{max}$, of fluorescent-based organic light emitting diodes (OLEDs) occurs when the probability that the injected carriers form excitons and the quantum yield for singlet emission are both unity. $\eta_{max}$ is then determined by (and identical to) the fraction, $f_s$ of injected electrons and holes (or negative and positive polarons, respectively) that pair to form emissive spin-singlet excitons, rather than nonemissive triplet excitons. If the process by which these excitons form were spin independent, then $\eta_{max}$ would be limited to 25\% based on spin-degeneracy. However, recent reports indicate that $\eta_{max}$ in OLEDs ranges between 22\% to 83\% \cite{rAlq3,rmeasurement1,rmeasurement2,nature,Friendnature,prl,TripletCIA,BaldoPRBrecent}. The exact value of $\eta_{max}$ and the reason for this variation, however, have remained controversial. Indeed, even the notion that $\eta_{max}$ can be larger than 25 \% is currently not universally accepted \cite{BaldoPRBrecent}.

\subsection{Overview of experimental results}

\begin{figure}[t]
\includegraphics[width=0.7\columnwidth]{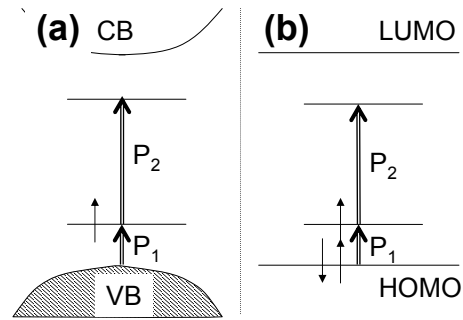}
\caption{\label{fig:FigPolaron} Models used to describe polaron levels and optical transitions depicted here for the positive polaron. (a) Electron-phonon (SSH) model. (b) Molecular orbital picture.}
\end{figure}

\begin{figure}
\includegraphics[width=\columnwidth]{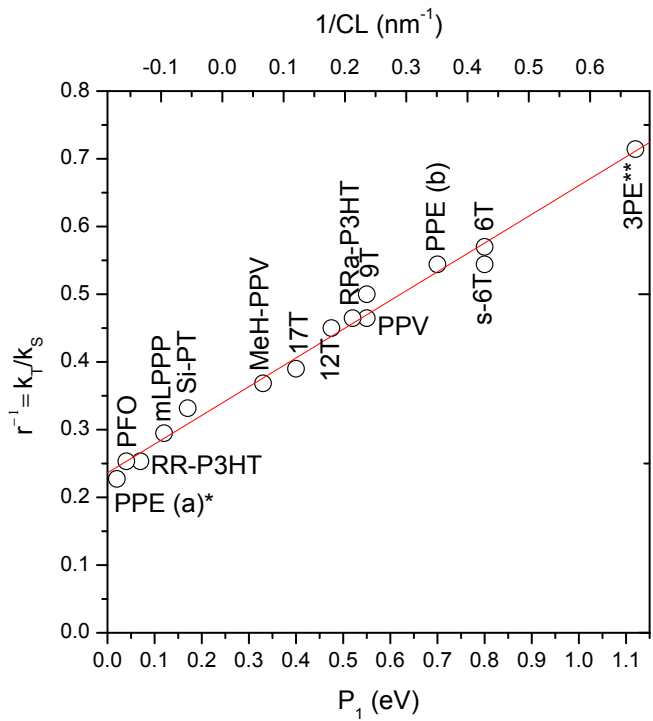}
\caption{\label{fig:rmat} Magnetic-resonance experimental data for the ratio $r^{-1}$=$k_T/k_S$ of spin-dependent exciton formation rates in various polymers and oligomers as a function of the peak photon energy of the $P_1$ transition (lower x-axis). $r^{-1}$ is also shown as a function of the inverse conjugation length 1/CL (upper x-axis), which was determined from $P_1$ (see text for discussion). The line through the data points is a linear fit. *The $P_1$ band of this polymer does not show a clear peak in the PA spectrum, the $P_1$ band extends to the longest wavelengths measured. **The length of this oligomer was calculated. In addition to the chemical names defined in the text, 3PE stands for the PPE trimer, PPE for poly(phenylene-ethynylene), Si-PT for silicon bridged polythiophene. For details consult original publications.}
\end{figure}

\begin{figure}[t]
\includegraphics[width=\columnwidth]{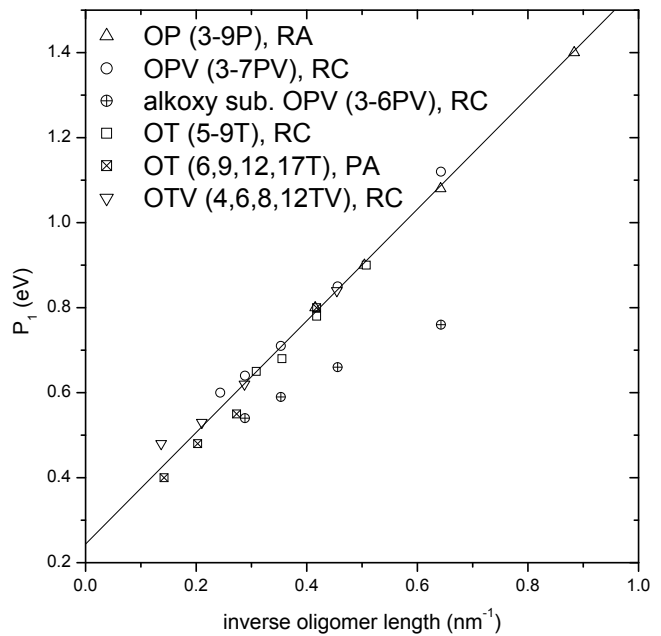}
\caption{\label{fig:oligomerdata} The peak photon energies of the $P_1$ polaron transition in a variety of oligomers, namely solutions of (unsubstituted) oligophenyls (OP, $\vartriangle$, radical anion (RA)), alkyl-substituted (AS) oligophenylene-vinylenes (OPV, $\circ$, radical cation (RC)), alkoxy-substituted OPV ($\oplus$, RC), end-capped oligothiophenes (OT, $\square$, RC), films of AS OT ($\boxtimes$, PA), AS oligothienylene-vinylenes (OTV, $\triangledown$, RC). The solid line is a fit to the data excluding $\oplus$.}
\end{figure}

Two entirely different experimental approaches have been employed to study spin-dependent exciton formation for OLEDs and thin films:

(i) Experiments \cite{rAlq3,rmeasurement1,rmeasurement2,Friendnature,TripletCIA,BaldoPRBrecent} that determine \emph{the singlet generation fraction} $f_S$ directly in "live" OLEDs. For fluorescent devices typically only the singlet emission can be measured, information on triplet density is missing and rather involved models have to be employed to obtain $f_S$ \cite{rmeasurement2}. Wilson et al. however, have recently shown \cite{Friendnature} that in OLEDs made from organic semiconductors that exhibit spin-orbit coupling, the strong intersystem crossing implies that both singlet and triplet emission (fluorescence and phosphorescence) can be simultaneously observed. This could be used to reliably determine $f_S$ by comparing the relative intensities of fluorescence to phosphorescence for optical excitation (where initially only singlet excitons are formed) with that for electrical excitation (where both singlet and triplet excitons are formed). Importantly, they found $f_S = 57$ \% for  devices made from a Platinum-containing polymer, but $f_S = 22$ \% for the corresponding monomer OLEDs. This suggests that exciton formation is spin-independent for the monomer, but that a spin-dependent formation process is effective in the polymer.

(ii) Experiments \cite{nature,prl,PRBRapid} that measure the ratio, $r=k_S/k_T$ of the spin-dependent \emph{exciton formation rates} for singlet and triplet excitons, respectively. Such experiments manipulate the spin state (using electron spin resonance techniques) of the pairing polarons, and measure the effect on exciton formation rates. These experiments consider photogenerated polarons in the film and use the fact that antiparallel spin polaron pairs can either form singlet or triplet excitons, whereas parallel spin pairs can only form triplets. These optically detected magnetic resonance (ODMR) techniques are modulation experiments where the resonant $\mu$-wave field is periodically turned on and off. Since the experiment is performed at low temperature, spin-alignment is conserved during the half-wave with $\mu$-wave field off, and polaron recombination/exciton formation obeys spin-statistics. However, during the half-wave with $\mu$-wave field on, spin-1/2 resonance leads to rapid spin-flips of the recombining polarons. Spin alignment is therefore not conserved, and each pair may choose whether to form singlet or triplet exciton. It can easily be shown \cite{nature,prl} that this leads to enhanced formation of the exciton with larger formation rate (leading to positive ODMR signal), at the expense of the more slowly forming exciton (that gives negative ODMR). In addition, the overall polaron recombination rate is enhanced, since the fast channel becomes allowed for all polaron pairs. Therefore changes occur in the photoinduced absorption (PA) from the triplet state, as well as the fluorescence from the singlet state upon magnetic resonance. In particular, from the $\mu$-wave induced change in PA of the polaron pairs, $r = k_S/k_T$ could be determined \cite{nature}.

\subsection{Polarons in $\pi$-conjugated semiconductors}

It is well-known that chemical doping or electrical charge injection results in the formation of polarons in $\pi$-conjugated semiconductors \cite{HeegerRevMod}. Fig. \ref{fig:FigPolaron} shows a comparison between different models that have been used for describing polarons in $\pi$-conjugated semiconductors. Panel (a) depicts the electron-phonon (e-p) or Su-Schrieffer-Heeger (SSH) model \cite{HeegerRevMod,Fesser1,Fesser2}. It predicts that the e-p coupling causes a gap between valence and conduction band.  In the singly charged system two localized polaron levels appear inside the gap. Experimentally one finds two optical transitions \cite{ValyPolaron} that are interpreted as the $P_1$ and $P_2$ transitions. Panel (b) depicts the molecular orbital picture where HOMO and LUMO are the highest occupied and lowest unoccupied molecular orbitals, respectively.

The most generic model for polarons is the molecular crystal or Holstein model \cite{Fesser1}; it is however not expected to be applicable in a quantitative way to $\pi$-conjugated polymers. This model yields for the "$P_1$" transition \cite{Fesser1}:

\begin{equation}
P_1=\left(\frac{A^2}{2 M\omega_E^2}\right)^2\frac{1}{W}
\end{equation}

Here A quantifies the e-p coupling strength (e.g. in eV/\AA), M is the ionic mass and $\omega_E$ is the Einstein phonon frequency, W is the band width before inclusion of e-p coupling. The term in the bracket is the energy, V associated with the e-p coupling \cite{P1Paper}. Furthermore, the following two equations hold for the polaron binding energy, $E_{b,polaron}$, and the deformation or relaxation energy $E_{relax,polaron}$, respectively.

\begin{eqnarray}
E_{b,polaron} & = & \frac{1}{3} P_1 \label{Eb} \\
E_{relax,polaron} & = & \frac{2}{3} P_1 \label{Erelax}
\end{eqnarray}

$\Pi$-conjugated oligomers are often used as model compounds instead of $\pi$-conjugated polymers because they can be obtained with a well-defined chemical structure. Although the molecular weight of polymers is typically much larger than that of oligomers, nevertheless it is established that the polymer should be viewed as a string of effectively independent segments, separated by chemical or physical defects. The length of these segments is called the conjugation-length (CL). Several recent calculations found that the energy associated with the e-p coupling decreases with increasing oligomer size. In particular, Devos and Lannoo found that $V=const/N$ in a more or less universal manner in acenes and fullerenes with various numbers, N of $\pi$-bonds \cite{P1Paper}; and Shuai et al. \cite{Shuai:1994} found that $E_{b,polaron}$ in oligophenyls (OP) decreases with the number of phenyl rings in a similar manner. We may therefore state that, in average, that effect of e-p coupling significantly decreases with increasing oligomer length.

\subsection{Results of magnetic resonance experiments}

Using ODMR, it was found that r is a monotonously increasing function of the conjugation-length (CL), and, by extrapolation, that $r \approx 1$ for small molecules and monomers \cite{prl}. Electroluminescence \cite{Friendnature} and magnetic resonance experiment therefore lead to the same qualitative conclusions, namely that exciton formation is spin-independent for the monomer, but that a spin-dependent process is effective in the polymer. Fig.~\ref{fig:rmat} shows the ratio $r^{-1}$=$k_T/k_S$ of spin-dependent exciton formation rates in various polymers and oligomers as a function of the peak photon energy of the $P_1$ transition (lower x-axis) obtained by ODMR spectroscopy \cite{prl}. In the original publication the experimental dependence $r(P_1)$ was however reinterpreted in terms of the CL of the polymer films studied. This was possible, because it is known \cite{polaronpaper} that there exists a (almost) universal relationship between $P_1$ and the material's CL (see Fig.~\ref{fig:oligomerdata}). $r^{-1}$ is therefore also shown as a function of 1/CL (upper x-axis), which was determined from $P_1$.

\subsection{Recent theoretical results}

\begin{figure}
\includegraphics[width=\columnwidth]{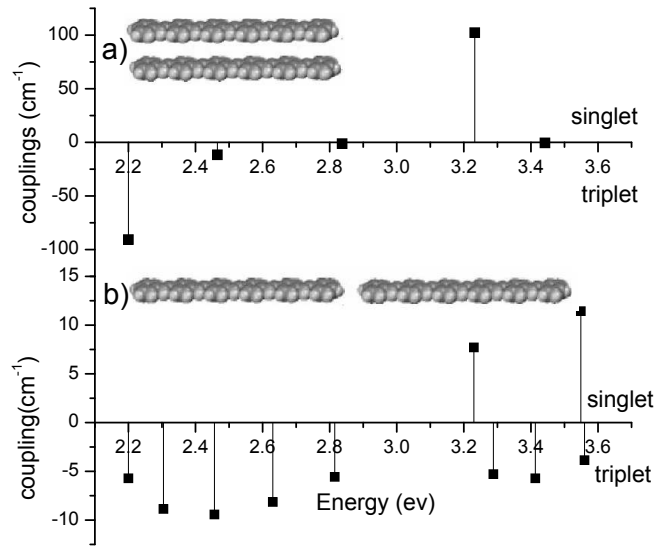}
\caption{\label{fig:ShuaiResults} Charge-recombination electronic couplings into singlet and triplet excited states in 6PV. Panel a) cofacial arrangement, panel b) head-to-tail arrangement. The charge-transfer state occurs at $\approx$ 3.7eV. The data points were taken from Ref. \cite{Beljonne:2004}.}
\end{figure}

Primary excitations in these materials are generally believed to be excitons with a binding energy in excess of kT, where k is the Boltzmann constant and T the temperature. As a result of electron-phonon and electron-electron interactions, the lowest singlet ($S_1$) and triplet ($T_1$) excitons posses both different energies (the $S_1-T_1$ energy difference; the exchange energy K for the lowest excitations has been measured \cite{Anna3} to be $K = 0.7 eV$ in a variety of conjugated polymers) and different spatial wavefunctions (with $T_1$ displaying a more spatially confined character).

Our model is based on recent work of Beljonne et al. \cite{Beljonne:2004}. They developed a theoretical model to describe intermolecular charge recombination in conjugated materials. In their treatment, Beljonne et al. found it necessary to consider two configurations of the two polymer chains involved in the exciton formation/polaron recombination process, namely cofacial and head-to-tail (see Fig.~\ref{fig:ShuaiResults}). They found that in the cofacial arrangement, by far the largest matrix elements are calculated for the lowest singlet $S_1$ and triplet $T_1$ excited states (see Fig.~\ref{fig:ShuaiResults} a). Similar results were obtained also by Tandon et al. \cite{Tandon:2003}. The situation is quite different for the head-to-tail configurations, where a number of different singlet and triplet excited states show significant electronic couplings to the polaron pair states (see Fig.~\ref{fig:ShuaiResults} b). The following picture has therefore emerged: Based on the electronic coupling, polaron recombination is a direct transition predominantly to the lower lying exciton states. Such transitions, however, have to pay a high price, since the multi-phonon emission necessary to conserve energy has very low probability for reasonable values of the Huang-Rhys parameter. Therefore, unless the two chains are in exact cofacial arrangement, the exciton states formed with highest probability may not be the lowest exciton states. We may therefore adopt a picture where exciton formation has highest probability for an intermediate exciton state.

\subsection{Theoretical approach}

In the case of charge-recombination (CR) processes, the semiclassical expression for the CR rate writes, within the Franck-Condon approximation, as \cite{Beljonne:2004}:

\begin{eqnarray}
k_{CR}=\frac{2\pi}{\hbar} \left | \left < \psi_i \left | W \right | \psi_f \right > \right |^2 \left (\frac{1}{4 \pi \lambda_S k T} \right )^{1/2} \times \\
\times \sum_\nu F_{0\nu} exp \left (-\frac{\left ( \Delta G^0+\lambda_S+\nu \hbar \omega_{ph} \right )^2}{4\lambda_S k T} \right ) \nonumber
\end{eqnarray}

Here $|\psi_i>$ and $|\psi_f>$ are the wavefunctions of the initial and final states, respectively, W is the perturbation, $\hbar \omega_{ph}$ is the energy of the most strongly coupled (optical) phonon, $F_{0\nu}$ is the Franck-Condon factor of the transition with zero and $\nu$ phonons in the initial and final states, respectively, $G^0$ is the difference in free energy between intial and final state, and $\lambda_S$ is the (external) reorganization energy. Next we approximate the above expression, since the largest contribution will result from transitions that conserve energy, namely those for which $\Delta G^0+\lambda_S+\nu \hbar \omega_{ph} \approx 0$. This can always be achieved, at least approximately, through emission of a number $\nu_E$ of phonons:

\begin{equation}
\nu_E=\frac{\Delta G^0+\lambda_S}{\hbar \omega_{ph}}
\end{equation}

We therefore obtain for $k_{CR}$:

\begin{equation} \label{Eq:kCR}
k_{CR} = \frac{2\pi}{\hbar} \left | \left < \psi_i \left | W \right | \psi_f \right > \right |^2 \left (\frac{1}{4 \pi \lambda_S k T} \right )^{1/2} F_{0\nu_E}
\end{equation}

The ratio, $r \equiv \frac{k_S}{k_T}$ is therefore given as:
\begin{equation} \label{Eq:r}
r=\frac{\sum_{n_S} k_{CR,S_{n_S}}}{\sum_{n_T} k_{CR,T_{n_T}}}
\end{equation}

where the sums extend over all singlet states and triplet states, respectively.

For the sake of simplicity, we now replace the sums in Eq.~\ref{Eq:r} by a single "effective" state $S$ and $T$, respectively. This effective state may be loosely identified with the intermediate exciton level discussed above, for which the exciton formation rate has maximum probability averaged over the ensemble of configurations. This step is justified in more detail in Appendix A. With this definition in mind we may write:

\begin{equation} \label{Eq:r2}
r=\frac{k_{CR,S}}{k_{CR,T}}= r_W \frac{F_{0\nu_S}}{F_{0\nu_T}}
\end{equation}

where the subscript S and T denote the "effective" singlet and triplet states, and $\nu_S$ and $\nu_T$ denote the number of phonons required for energy conservation to form the S and T states, respectively. The definition of $r_W$ follows by comparison of Eq.~\ref{Eq:r2} to Eq.~\ref{Eq:r}, and is essentially the ratio of the electronic matrix elements for singlet and triplet formation, respectively.

Next we discuss the calculation of $F_{0\nu}$, where we closely follow the treatment by Barford et al. \cite{Barford} Inter-molecular interconversion is an iso-energetic process which occurs from the lowest vibrational levels of the initial polaron pair state to the final, intra-molecular exciton states at the same energy as the initial level. In actuality, the exciton formation process involves two conformational transitions, namely the transition from the polaron lattice conformation to the exciton conformation in chain 1 and from polaron to groundstate conformation in chain 2. We therefore need to generalize our treatment in the following way:

\begin{eqnarray}
F_{0\nu_E} & = & \sum_{\nu_1 \nu_2} F^{(1)}_{0\nu_1} F^{(2)}_{0\nu_2} \delta (E_f-E_i) \\
& = & \sum_{\nu_1} F^{(1)}_{0\nu_1} F^{(2)}_{0(\nu_E-\nu_1)}
\end{eqnarray}

where $F^{(1)}_{0\nu_1}$ and $F^{(2)}_{0\nu_2}$ are the Franck-Condon factors associated with the vibrational wavefunction overlaps of chains 1 and 2, respectively.

\begin{figure}
\includegraphics[width=\columnwidth]{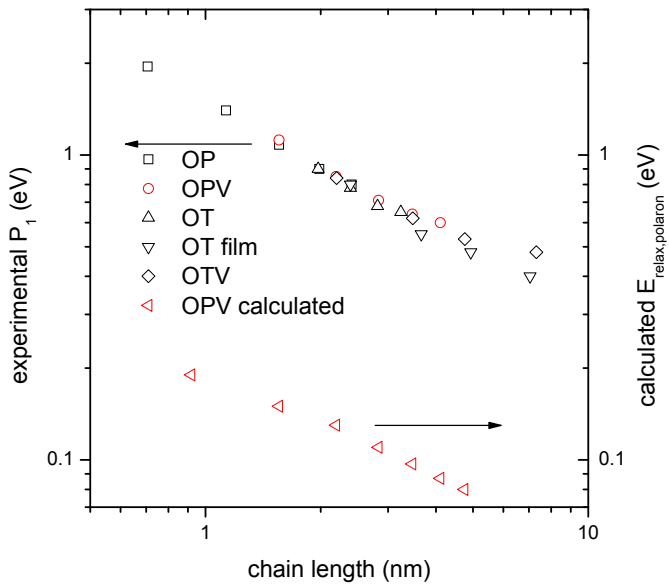}
\caption{\label{fig:Fig1} Experimental values for $P_1$ polaron transition energy in a variety of oligomers, namely solutions of (unsubstituted) oligophenyls (OP, radical anion (RA)), alkyl-substituted (AS) oligophenylene-vinylenes (OPV, radical cation (RC)), end-capped oligothiophenes (OT, RC), films of AS OT (measured by photoinduced absorption), AS oligothienylene-vinylenes (OTV, RC) together with calculated values for the polaron relaxation energy $E_{relax,polaron}$ in OPV. The experimental data for $P_1$ were taken from Ref, whereas the calculated values for $E_{b,polaron}$ were taken from Ref.}
\end{figure}

Next we want to write out the expression for the Franck-Condon overlaps using the displaced oscillator expression $F_{0\nu}=\left ( e^{-S} S^\nu \right )/\nu!$. A useful simplification to this expression arises by noting that the geometric distortions of the polarons and exciton polarons (namely the $1^1B_u$ or $1^3B_u$ states) from the ground state structure are very similar. \cite{Barford:2001} Thus, the Huang-Rhys parameter, $S_1$ for the $1^1B_u$ and $1^3B_u$ states relative to the positive polaron is quite small. Therefore, to the lowest approximation, only $\nu_1=0$ has a non-negligible contribution, and $S_{total} \approx S_2 \equiv S_P$, where the P stands for polaron conformation. In the next higher approximation, we may also include the term $\nu_1=1$. As we show in Appendix B, if $S_1 \ll 1$ and $S_1 \ll S_P$, then we may combine the two Franck-Condon factors into a single one with Huang-Rhys parameter, $S=S_1+S_P$.

Therefore:

\begin{eqnarray}
r & = & r_W \left ( S_1+S_P \right )^{-(\nu_T-\nu_S)} \frac{\nu_T !}{\nu_S !} \label{Eq:final1} \\
& = & r_W \left ( S_1 + \frac{E_{relax,polaron}}{\hbar \omega_{ph}} \right )^{-\frac{\Delta_{S/T}}{\hbar \omega_{ph}}} \frac{\nu_T !}{\nu_S !} \label{Eq:finalB}
\end{eqnarray}

Eqs.~\ref{Eq:final1} and \ref{Eq:finalB} are the final result of our model. 

\section{Comparison between experimental results and model}

Guided by the magnetic resonance experimental data shown in Fig.~\ref{fig:rmat}, we now study the CL dependence of Eq.~\ref{Eq:final1}. The CL-dependence of $r_W$ was studied previously by Beljonne et al. \cite{Beljonne:2004} They found that $r_W$ depends only weakly on the CL, in particular this dependence is not strong enough to account for the material dependence of r shown in Fig.~\ref{fig:rmat}. We must therefore study the CL dependence of the phonon emission term in Eq.~\ref{Eq:final1}.

There is an obvious connection between our model and the experimental magnetic resonance results. These results (see Fig.~\ref{fig:rmat}) are given as a function of $P_1$, whereas our model can be written in terms of $E_{relax,polaron}$. The Holstein model relates the two quantities through Eq.~\ref{Erelax}. A similar relationship should therefore hold between $P_1$ and $E_{relax,polaron}$ in $\pi$-conjugated polymers. However, the Holstein model is formulated for the infinite system, and does not consider finite size effects. It is less than obvious that $P_1 \propto E_{relax,polaron}$ holds also in oligomers. In particular, since we are interested in phonon emission, we need to distill finite size effects on the e-p coupling energy from finite size effects on the electronic energies (quantum confinement energy). Finite size effects on the e-p coupling energy have been studied by Devos and Lannoo, who found that $V=const/N$ in a more or less universal manner in acenes and fullerenes with various numbers, N of $\pi$-bonds \cite{P1Paper}. We therefore expect that $E_{relax,polaron}$ decreases strongly as the oligomer size increases.

\subsection{The conjugation-length dependence of the polaron relaxation energy}

To the best of our knowledge a direct measurement of $E_{relax,polaron}$ has not yet been performed in $pi$-conjugated polymers and oligomers. However, Bredas and coworkers have calculated \cite{Shuai:1994} $E_{relax,polaron}$ for phenyl-capped PPV oligomers as a function of the number of rings N against which we can compare our experimental results for $P_1$. Their results are shown in Fig.~\ref{fig:Fig1}. Fig.~\ref{fig:Fig1} also shows a plot of the $P_1$ polaron transition energy in a variety of oligomers. It is seen in Fig.~\ref{fig:Fig1} that both experimental data for $P_1$ and theoretical calculation of $E_{relax,polaron}$ follow a very similar dependence on CL, as a matter of fact $P_1 \propto E_{relax,polaron}$ to a high degree of accuracy. This is strong evidence for the notion that $P_1$ is indeed a measure of $E_{relax,polaron}$. Therefore both experimental and theoretical data find the that $E_{relax,polaron}$ monotonously decreases with increasing CL. In particular (see Fig.~\ref{fig:oligomerdata}),

\begin{equation}
E_{relax,polaron}=E_{relax,polaron,\infty}+k \times \frac{1}{CL}
\end{equation}

Through comparison of the data shown in Figs.~\ref{fig:rmat} and \ref{fig:Fig1} it is therefore evident that the experimental results can be brought in agreement with Eq.~\ref{Eq:final1} only if $\nu_T-\nu_S =1$, therefore

\begin{eqnarray}
r^{-1} & = & \frac{S_1 + \frac{E_{b,polaron}}{\hbar \omega_{ph}}}{r_W \nu_T} \label{Eq:final2} \\
\end{eqnarray}

Importantly, since the experimental relationship $P_1 (CL)$ is universal, then it follows naturally that $r (CL)$ is universal.

\subsection{The conjugation-length dependence of the optical Huang-Rhys parameter}

\begin{figure}
\includegraphics[width=\columnwidth]{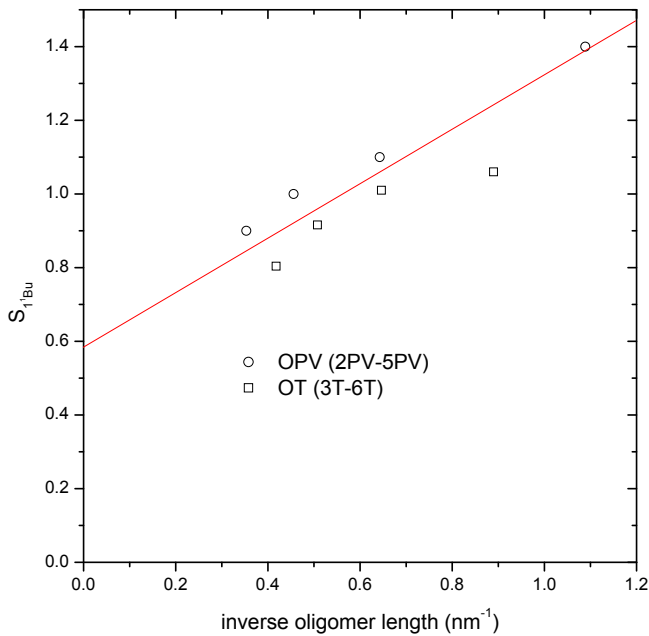}
\caption{\label{fig:S1Bu} Experimental values for the optical Huang-Rhys parameter, $S_{1^1B_u}$, in a series of OPV and OT.}
\end{figure}

\begin{figure}
\includegraphics[width=\columnwidth]{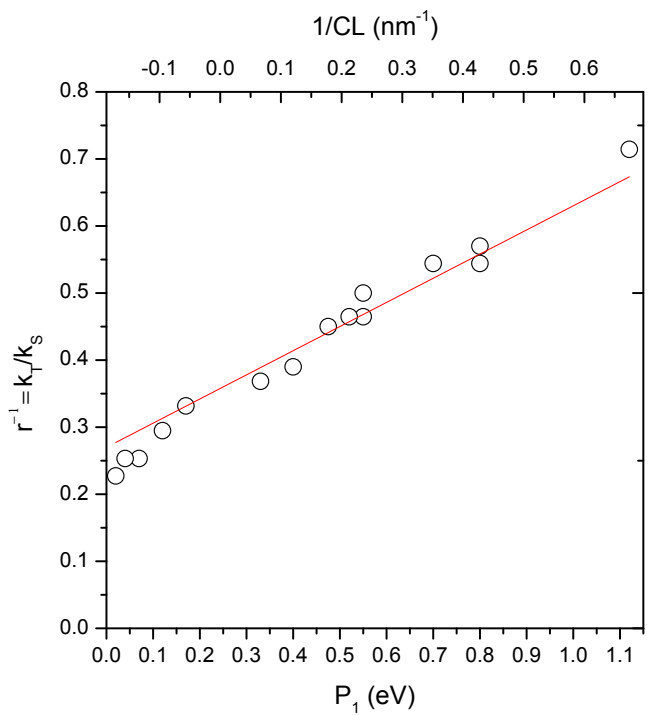}
\caption{\label{fig:S1BuFit} A fit of the experimental data for $r^{-1}$ using our multi-phonon emission model and together with the experimentally determined optical Huang-Rhys parameter, $S_{1^1B_u}$.}
\end{figure}

Our model is based on the assertion that the Huang-Rhys parameter is substantially smaller in polymers than oligomers. Thus far, we have presented three justifications for this statement:

\begin{enumerate}
    \item The optical polaron transition energy $P_1$ strongly scales with CL, specifically as $P_1=P_{1,\infty}+k_{P_1} \times CL$. Furthermore this scaling is universal.
    \item A very similar scaling law was found by calculations of $E_{relax,polaron}$ vs. CL in OPV performed by Bredas and coworkers.
    \item The electron-phonon energy, V was found to scale as $V \propto 1/N$, where N is the number of $\pi$-atoms in a large class of conjugated compounds, largely independent of the specific chemical structure.
\end{enumerate}

Here we now want to add another set of data to justify the main assertion on which our model is built. The lattice relaxation of the polaron state is expected to be similar to that of the lowest singlet exciton state $1^1B_u$. The corresponding Huang-Rhys parameter, $S_{1^1B_u}$ can be easily measured by optical absorption and emission. We may therefore test our assertion by determining $S_{1B_u}$ from absorption and emission spectra of oligomers.

Fig.\ref{fig:S1Bu} shows experimental values for the optical Huang-Rhys parameter, $S_{1^1B_u}$, in a series of OPV and OT. The experimental data were taken from Ref.~\cite{Cornil:1997}  and Ref.~\cite{Yang:1998}. It is seen that $S_{1^1B_u}$ decreases with increasing CL, namely $S_{1^1B_u}=S_{1^1B_u,\infty} + k \times \frac{1}{CL}$ in agreement with our expectation based on $E_{relax,polaron}$.

Fig.~\ref{fig:S1BuFit} shows the fit of the experimental data for $r^{-1}$ using our multi-phonon emission model together with the experimentally determined optical Huang-Rhys parameter, $S_{1^1B_u}$. Specifically we used the fit function

\begin{equation}
r^{-1} = \frac{S_{1^1B_u}}{r_W \nu_T} \label{Eq:final3}
\end{equation}

It is seen that excellent agreement is achieved.

\section{Summary}

We developed a model of spin-dependent exciton formation in OLEDs. We calculated $r=k_S/k_T$ based on a multi-phonon emission process. The resulting equation for r therefore strongly depends on the Franck-Condon overlap integrals that are parameterized by the Huang-Rhys parameter. Guided by recent magnetic resonance experiments, we studied the dependence of Huang-Rhys parameter on the conjugation-length. We used two different approaches:

\begin{enumerate}
    \item We relate the Huang-Rhys parameter to the polaron relaxation energy. Then we relate $E_{relax,polaron}$ to the optical polaron transition $P_1$, which has been measured experimentally in many oligomers. This procedure leads us to conclude that $S=S_\infty + k \times \frac{1}{CL}$.
    \item We relate the Huang-Rhys parameter in our model to the Huang-Rhys parameter that can be measured by photoluminescence spectroscopy. Such data are available in OPV and OT. This procedure also leads us to conclude that $S=S_\infty + k \times \frac{1}{CL}$.
\end{enumerate}    

Our work therefore leads to the following picture of exciton formation: Since the triplet exciton states lie lower in energy than singlets, more phonons must be omitted (required by energy conservation) for triplet formation than singlet formation. Since polymers have a small Huang-Rhys factor, then the emission of many phonons is unlikely, thus favoring singlet formation. In short oligomers, however, the Huang-Rhys factor is quite large, phonons are emitted easily, and singlet and triplet formation both become likely.

\section{Appendix A}

The definition of a single effective state is certainly appropriate in the cofacial configuration, since only the lowest exciton states contribute to the exciton formation process. In the head-to-tail configuration, however a whole series of states contribute. Fig.~\ref{fig:ShuaiResults} b) shows that the electronic couplings are roughly equal in all states that contribute, such that we may write:

\begin{eqnarray}
\sum_n\frac{2\pi}{\hbar}\left | \left < \psi_i \left | W \right | \psi_{f,n} \right > \right |^2 \left (\frac{1}{4 \pi \lambda_S k T} \right )^{1/2} F_{0\nu_{E,n}} \\
\approx \frac{2\pi}{\hbar} \overline{\left | \left < \psi_i \left | W \right | \psi_{f,n} \right > \right |^2} \left (\frac{1}{4 \pi \lambda_S k T} \right )^{1/2} \sum_n F_{0\nu_{E,n}} \nonumber
\end{eqnarray}

Therefore the effective state in the head-to-tail arrangement has an electronic matrix-element $\overline{\left < \psi_i \left | W \right | \psi_{f,n} \right >}$, and Franck-Condon Factor equal to $\sum_n F_{0\nu_{E,n}}$, and has a binding energy $\nu_E \hbar \omega_{ph}$, such that $F_{0\nu_E}=\sum_n F_{0\nu_{E,n}}$.

\section{Appendix B}

Including only the $\nu_1=0$ and $\nu_1=1$ terms we then obtain

\begin{eqnarray}
F_{0\nu_E} & = & F^{(1)}_{00} F^{(2)}_{0\nu_E} + F^{(1)}_{01} F^{(2)}_{0(\nu_E-1)} \\
& = & \frac{e^{-S_1}e^{-S_P}S_P^{\nu_E}}{\nu_E!}+\frac{e^{-S_1}S_1e^{-S_P}S_P^{\nu_E-1}}{(\nu_E-1)!} \\
& = & \frac{e^{-(S_1+S_P)} \left ( S_P^{\nu_E} + \nu_E S_1S_P^{\nu_E} \right )}{\nu_E!}
\end{eqnarray}

On the other hand, the Franck-Condon factor, $F_{0\nu_{max}}$ for a (hypothetical) Huang-Rhys parameter, $S=S_1+S_P$ equals

\begin{equation}
F_{0\nu_E} = \frac{e^{-(S_1+S_P)} \left ( S_1+S_P \right )^{\nu_E}}{\nu_E!}
\end{equation}

We see that if $S_1 \ll S_P$, the two expressions are equal to first order.

\bibliography{Spaper}

\end{document}